\documentstyle[12pt,epsf]{article}
\def\lsim{>\kern-2.5ex\lower0.85ex\hbox{$\sim$}\ }
\def\rsim{>\kern-2.5ex\lower0.85ex\hbox{$\sim$}\ }

\begin{document}

\baselineskip 16pt

\vspace{.25in}

\centerline{\large\bf NEUTRINO SCATTERING IN A MAGNETIC FIELD}

\vspace{.25in}

\centerline{\large K.S. McFarland$^1$, A.C. Melissinos$^1$}
\centerline{\large N.V. Mikheev$^2$ and W.W. Repko$^3$}

\bigskip

\centerline{\small{ $^1$Department of Physics and Astronomy,
University of Rochester, Rochester, NY 14627, USA}}

\centerline{\small{$^2$Department of Physics, Yaroslavl' State
University, Yaroslavl', 150000 Russia}}

\centerline{\small{$^3$Department of Physics and Astronomy,
Michigan State University, East Lansing, MI 48824, USA}}

\bigskip
\centerline{\today}

\vspace{.25in}

\begin{abstract} Motivated by the evidence for a finite neutrino mass we examine
 anew the interaction of neutrinos in a
magnetic field.  We present the rate for radiative scattering for
both massless and massive neutrinos in the standard model and give
the corresponding numerical estimates.  We also consider the
effects arising from a possible neutrino magnetic moment.
\end{abstract}

\vspace{.50in}

\noindent{\large\bf 1. Introduction}
\medskip

Neutrino flavor oscillations, for which there now appears
significant experimental evidence, have established that at least
some neutrino has non-zero mass.  However the sensitivity to mass
in vacuum flavor oscillations only enters as $\Delta m^2$, and
thus one is unable to establish the neutrino's mass.
Electromagnetic interactions of neutrinos, however, often have
direct sensitivity to the neutrino mass.  In this paper we
evaluate a number of electromagnetic processes that could result
when a high energy neutrino enters a laboratory magnetic field and
consider the practicality of observing such effects.

We first consider the radiative scattering of high energy
neutrinos in a magnetic field.  Such scattering arises from the
coupling to the field of the virtual charged particles in the loop
diagrams shown in Fig.1.  Radiative scattering can also be viewed
as the back-scattering of the virtual photons in the magnetic
field, off the high energy neutrinos.

Neutrino photon scattering has been considered early on, and it
was shown by Gell-Mann [1], that for a local interaction and
massless neutrinos, $\nu +\gamma \to \nu + \gamma$ is forbidden by
angular momentum conservation [2].  This restriction does not
apply for virtual photons, massive neutrinos or to higher order in
$(m_\ell/m_W)^2$. Dicus and collaborators [3-6] have given cross
sections for both massless and massive neutrino photon scattering.
Neutrino scattering in a magnetic field has been considered in
detail by Mikheev and collaborators [7-10], as well as by other
authors [11,12].

For high energy neutrinos, the radiated photons are energetic and
the distribution is peaked in the forward direction.  The
kinematics can be visualized in terms of the 3-momentum $\vec{q}$
of the virtual photons.  We take $\vec{q}$ to be directed opposite
to the neutrino momentum and $q^0$ = 0.  With $E,\vec{p}$ and
$\vec{E}',\vec{p}\ '$ the incoming and outgoing neutrino
4-vectors, $m_\nu$ the neutrino mass and $\omega'$ the scattered
photon energy,

\begin{equation}
\omega' = \frac{E}{1+(m_\nu/2\gamma|q|)(1+\gamma^2\theta^2)}
\end{equation}

\noindent Here $\gamma = E_\nu/m_\nu$ and $\theta$ is the
scattering angle measured from the forward direction.

In the following section we present the scattering rate as
calculated in the standard model.  This is followed by numerical
estimates of expected rates for operating and planned
accelerators.  We conclude with a discussion of effects that would
manifest themselves should neutrinos possess a magnetic moment.

\vspace{.25in}

\noindent{\large\bf{2. Scattering Rate in the Standard Model}}

\medskip

The relevant amplitudes are shown in Fig.1.  For small momentum
transfers, $s/m_W^2 \ll 1$, the boson propagators can be
contracted to the low-energy 4-fermion interaction.  It is also
evident from the graphs, that the dominant contribution will come
from the lowest mass fermion, in the loop the electron.  The
$W$-exchange graph allows for mixing of the neutrino flavor
eigenstates, specified by a unitary matrix $U_{\ell\alpha}$, where
$\ell, \alpha$ are the flavor, mass, eigenstate indices. The
external field appears in the neutrino rest-frame as time
independent crossed electric and magnetic fields.  The exact
solution of the wave equation in these fields is used for the
lepton propagator [7] instead of the virtual photon coupling to
the external field.

The scattering rate (probability per unit time) is expressed in
terms of the invariant $\chi_e$  defined through

\begin{equation}
\chi_e^2 = \frac{e^2(p_\nu FFp_\nu)}{m_e^6} =
\left(\frac{E_\nu}{m_e}\right)^2\left(\frac{B}{B_e}\right)^2 =
E_\nu^2\left(\frac{eB}{m_e^3}\right)^2
\end{equation}
where $B_e = 4.41 \times 10^9$ T is the Schwinger critical field
[13], and $m_e$ the electron mass.  Three different cases can be
identified.  First, for massless neutrinos, $m_\nu = 0$, we find

\begin{equation}
\Gamma = \frac{\alpha}{16\pi} \frac{(G_Fm_e^2)^2}{(2\pi)^3}\
\left[\frac{5}{216}\right]\ \frac{m_e^2\chi_e^6}{E_\nu}
\end{equation}

In the second case the neutrino is assumed massive, $m_\nu \neq
0$.  In the absence of mixing the scattering rate is given by

\begin{equation}
\Gamma = \frac{\alpha}{16\pi}\ \frac{(G_F m_e^2)^2} {(2\pi)^3}
 \left[\frac{185\cdot 11}{9\cdot(30)^3}\right]
\frac{m_\nu^2 \chi_e^4}{E_\nu}
\end{equation}

\noindent In the presence of mixing but if the incident mass
eignestate, $\alpha$, is not changed by the scattering Eq.(4)
remains valid but is modified by a factor (given here for 3-flavor
mixing)

$$ \left|U^*_{e\alpha} U_{e\alpha} -\left(U^*_{\mu\alpha}
U_{\mu\alpha} + U^*_{\tau\alpha} U_{\tau\alpha} \right)\right|^2
\eqno(4a) $$

\noindent This factor arises from the interference of the $W$ and
$Z_0$ contributions.

The third cases arises in the presence of mixing but when a
massive neutrino $\alpha$, decays radiatively to a lighter
neutrino $\beta$.  Assuming that $m_{\nu\alpha} \gg m_{\nu\beta}$,
the rate for the process $\nu_\alpha \to \nu_\beta + \gamma$ is
now

\begin{equation}
\Gamma = \frac{\alpha}{16\pi} \frac{(G_F m_e^2)^2}{(2\pi)^3}
\left[\frac{1}{108}\right]\left(\frac{m_{\nu\alpha}}{m_e}\right)^2
\frac{m_{\nu\alpha}^2\chi_e^2}{E_\nu}\left|\frac{}{}U_{e\alpha}^*U_{e\beta}\right|^2
\end{equation}

\noindent Of course in an experiment the incident as well as the
detected neutrinos are usually flavor eigenstate.  Thus to use
Eq.5 or Eq.4a one has to first project the flavor eigenstate onto
it mass eignestates.

The above results remain valid as long as $\chi_e <1$ which is
always the case even at the highest neutrino (beam) energies and
for magnetic fields that can be achieved in the laboratory.
Results pertaining to larger values of $\chi_e$ such as can be
reached in the very strong fields of astrophysical environments or
for extremely high energy cosmic ray neutrinos are discussed in
[9,14].

We also show the neutrino photon-scattering cross section for two
special cases: (a) Massless neutrino but the initial photon is
virtual. In this case we set (in the lab frame) $q^\mu =
\{0,0,0,-q\},\ p^\mu = \{E_\nu, 0, 0, p_\nu\}$ and find

\begin{equation}
\sigma = \frac{\alpha^2}{16\pi}\ \frac{G^2_Fm_e^2}{(4\pi)^2}\
\left[\frac{5}{54}\right]\ \frac{s}{4m_e^2}\
\left(\frac{q^2}{m_e^2}\right)^2
\end{equation}
Here $s = 2E_\nu|q|$ is the square of the cm energy and $q^2 =
-q_\mu q^\mu$.

\noindent (b) When the incoming photon is real but the neutrino is
massive. The cross section for this process has been discussed in
detail in ref. [6].  When $s\gg 4m_e^2$ the cross section tends to
a constant value

\begin{equation}
\sigma \simeq \frac{\alpha^2}{16\pi}\
\left(\frac{G_F^2m_\nu^2}{4\pi^2}\right)\
\left(\frac{s-m_\nu^2}{s}\right)^2
\end{equation}
Eqs.6,7 can be used in conjunction with the virtual photon
formalism alluded to previously to obtain estimates of the
scattering rates in a magnetic field.

\vspace{.25in}

\noindent{\large\bf{3. Numerical Estimates}}

\medskip

The range of variables entering eqs.(3-5) is restricted by the
experimental possibilities.  We will therefore consider only the
following values (in Gaussian units)

$$\begin{array}{llllllll} B & = & 2.2\ {\rm T}\ \simeq  2.2 \times
693 &{\rm eV}^2\qquad \\ L & = & 10\ {\rm m} \simeq 5 \times 10^7
& {\rm eV}^{-1}\\ E_\nu & =& 50\ {\rm GeV}\end{array}$$

\noindent It follows that $\chi_e \simeq 0.5 \times 10^{-4}$ and
for the purposes of this estimate we have set
$U_{e\alpha}^*U_{e\beta} = 1$ in Eq.5. The resulting probabilities
for radiative scattering per incident neutrino are shown in Fig.2,
as a function of neutrino mass.  It is important to note that
Eqs.(3-5) have been obtained by using the amplitude corresponding
to only one of the three possible cases.  Thus for a given
neutrino mass the probability for radiative scattering is given by
the dominant contribution: see for instance Fig.2.

We note that for $m_\nu \rsim$ 100 eV the dominant process is the
radiative decay catalyzed by the presence of the magnetic field.
The probabilities shown in Fig.2 are to be compared to the
available integrated fluxes of high energy neutrinos.  The MINOS
beam at Fermilab will soon deliver 10$^{18}$, 50 GeV neutrinos per
year, mainly $\nu_\mu$.  A future 50 GeV neutrino factory could
deliver, under optimal conditions 10$^{21}$ neutrinos in one year
at $E_\nu \simeq$ 20 GeV. In either case these fluxes are too low
to lead to observable effects unless $m_\nu > 100$ MeV.

\vspace{.25in}

\noindent{\large\bf 4. Magnetic Moment Interactions}

\medskip

A Majorana neutrino or a massless Dirac neutrino can have no
magnetic moment.  A massive Dirac neutrino has a magnetic moment
which arises in the standard model from loop corrections. To
leading order in $(m_\ell/m_W)^2$ it has the value [15]

\begin{equation}
\mu_\nu = \frac{3e G_F\ m_\nu}{8\pi^2\sqrt{2}} = 3.2 \times
10^{-19} \left(\frac{m_\nu}{1\ {\rm eV}}\right) \ \mu_0
\end{equation}
where $\mu_0$ is the Bohr magneton, $\mu_0 = 5.79 \times 10^{-11}$
MeV/T.

The accepted limits on possible magnetic moments as given by the
PDG [16] are

$$
\begin{array}{lccl} \mu_{\nu 1} \lsim 10^{-10}\ \mu_0 & \qquad & \qquad & {\rm
electron \ neutrino}\\ \mu_{\nu 2} \lsim 10^{-9}\ \mu_0 & & & {\rm
muon \ neutrino}\\ \mu_{\nu 3} \lsim 10^{-6}\ \mu_0 & & & \tau
{\rm -neutrino}
\end{array}
$$

\noindent These limits are obtained most directly from the shape
and rate of the spectrum in $\nu e$ and $\bar{\nu}e$ scattering
(see for instance Ahrens et al [17]).  There are also
astrophysical limits based on the cooling rate of stars and on the
observation of the neutrino burst from SN1987A.  Such limits are
in the range of $(10^{-10}$ to $10^{-12})\mu_0$ for all three
neutrino mass eigenstates. We discuss three possible
manifestations of a magnetic moment interaction in a magnetic
field.  For laboratory field strengths these processes are far
from reaching the value predicted by Eq.8, nor can they improve on
the existing limits listed previously.

The simplest interaction is the precession of the spin vector
which modifies the helicity state of the neutrino and thus alters
its weak interaction rate.  Spin rotation is due to the different
time evolution of the two spin states projected onto the magnetic
field; the neutrino momentum vector is assumed to be perpendicular
to the magnetic field. The rotation angle is
\begin{equation}
\theta = 2\mu_\nu BL
\end{equation}
where $L$ is the length of the field.  The fractional change in
the weak cross-section is then

\begin{eqnarray}
\Delta\sigma/\sigma \simeq 1 -\cos^2\theta\nonumber\\
\noalign{\hbox{ and for small rotation angles}}
\Delta\sigma/\sigma \simeq 4\mu_\nu^2 B^2L^2
\end{eqnarray}

\noindent For $B=2$ T and $L=10$ m a change $\Delta\sigma/\sigma
\lsim 1\%$ would correspond to a limit $\mu_\nu/\mu_0 \lsim
10^{-5}$.

If we allow for mixing, different mass eigenstates would have
different magnetic moments. In the presence of flavor mixing among
magnetic moment eigenstates, the presence of an axial magnetic
field would lift the degeneracy between these eigenstates, and
potentially result in field-induced neutrino flavor oscillations.
Analyzing this situation in the familiear two neutrino case where
we assume the mass and magnetic moment eigenstates to be
identical, as in the standard model, we find the flavor
oscillation probability:

$$P(\nu_\alpha \to \nu_\beta) =\sin^2 2\theta \sin^2 (\Delta
m_\nu^2 L/4E + \Delta\mu_\nu BL/2)$$

\noindent where $\theta$ is the mixing angle, $\Delta m_\nu^2 =
\left|m_\alpha^2 - m_\beta^2\right|\ {\rm and}\ \Delta\mu =
\left|\mu_\alpha -\mu_\beta\right|$. The second term in the time
evolution factor is typically smaller than the first term, except
in cases wehre the magnetic moment is anomalous, the field
strength is extreme or the neutrino is very energetic $(E_\nu =
10^{12}$ GeV when $(m_\alpha + m_\beta) \sim 0.1\ {\rm eV\ and}\ B
= 2\ {\rm T}).$

In a laboratory experiment, one can search for an anomalous
neutrino magnetic moment through this process.  For $B = 2$ T, $L
= 10$ m and if one can detect oscillations at the 10$^{-4}$ level,
then for maximal mixing the limit is $\Delta\mu/\mu_0 < 2 \times
10^{-6}$.

The presence of a magnetic moment would also lead to the emission
of a high energy photon by magnetic \lq\lq Compton" scattering as
shown in Fig.3. For the cross-section for this process involving
real or virtual photons and assuming that $s\gg q^2$, we find

\begin{equation}
\sigma = \frac{\pi\alpha^2}{8m_e^2}
\left(\frac{\mu_\nu}{\mu_0}\right)^4\ \frac{s}{m_e^2}
\end{equation}
where $s$ is the cm energy.  For a head-on collision with a real
photon of momentum $q, s = 4 E_\nu q$. Using the virtual photon
formalism [14] we can obtain the radiation rate in a magnetic
field $B$

\begin{equation}
\Gamma \simeq \frac{1}{8} \alpha\
\left(\frac{\mu_\nu}{\mu_0}\right)^4\ \frac{m_e^2}{E_\nu}\
\chi_e^2
\end{equation}
where $\chi_e$ is the invariant of Eq.3.  For $E_\nu$ = 50 GeV,
and $\chi_e = 0.5 \times 10^{-4} (B = 2.2\ {\rm T})$,  $L = 10\
{\rm  m}$ and $(\mu_\nu\mu_0) = 7 \times 10^{-5}$ the scattering
probability is $P \simeq 10^{-20}$. Given the available neutrino
fluxes this can be considered as the lowest limit of detection.

Finally we note that the presence of a transition magnetic moment
leads to radiative decay.  The invariant probability (inverse
lifetime in the neutrino rest frame) is [18]

\begin{equation}
\Gamma(\nu_\alpha \to \nu_\beta) =
\left(\frac{\mu_{\alpha\beta}}{\mu_0}\right)^2\ \mu_0^2
\frac{(\Delta m_{\alpha\beta}^2)^3}{8\pi m_e^3}
\end{equation}

\noindent For instance for $L=10$ m, $E_\nu$ = 50 GeV and $\Delta
m_{\alpha\beta}^2 = m_\alpha^2 = 1$ eV$^2$, the decay probability
is

\begin{equation}
P = 3.5 \times 10^{-18}(\mu_{\alpha\beta}/\mu_0)^2
\end{equation}

\noindent It has been proposed to use an external radio frequency
field to induce the transition [19].  As an example, using one of
the LEP rf cavities at nominal power levels the probability for a
(non-radative) transition is of order

\begin{equation}
P \sim 10^3(\mu_{\alpha\beta}/\mu_0)^2
\end{equation}
Experimentally, the transition leads to a change in the rate of
detection of a particular neutrino flavor.  This limits the
observable probability to $P\rsim 10^{-3}$, or\\
$(\mu_{\alpha\beta}/\mu_0)<10^{-3}$.

\vspace{.25in}

This work was supported in part under DOE grant DE-FG02-91ER40685;
The work of N.V.Mikheev was supported in part by the Russian
Foundation for Basic Research under the Grant No. 01-02-17334.

 \vfil\eject

 \centerline{\large\bf References}

\bigskip
\begin{enumerate}
\item M. Gell-Mann, Phys. Rev. Lett. \underline{6}, 70 (1960).
 \vspace{-7pt}
 \item C.N. Yang, Phys. Rev. \underline{77}, 242 (1950), L.D.
 Landau, Sov. Phys. Dokl. \underline{60}, 207 (1948).
 \vspace{-7pt}
\item D.A. Dicus and W.W. Repko, Phys. Rev. Lett. \underline{79},
569 (1997).
 \vspace{-7pt}
\item D.A. Dicus, C. Kao and W.W. Repko, Phys. Rev.
\underline{D59}, 013005 (1998).
 \vspace{-7pt}
\item D.A. Dicus and W.W. Repko, hep-ph/0003305.
 \vspace{-7pt}
\item D.A. Dicus, W.W. Repko and R. Vega, hep-ph/0006264.
 \vspace{-7pt}
\item LA. Vasilevskaya, A.A. Gvozdev, N.V. Mikheev, Yadern. Fizika
\underline{57}, 124 (1994), Physics of Atomic Nuclei
\underline{57}, 117 (1994).
 \vspace{-7pt}
\item A.V. Kuznetsov and N.V. Mikheev, Physics Letters
\underline{B299}, 367 (1993).
  \vspace{-7pt}
\item A.A. Gvozdev, N.V. Mikheev and L.A.
Vassilevskeya, Phys. Rev. \underline{D54}, 5674 (1996).
 \vspace{-7pt}
\item A.V. Kuznetsov and N.V. Mikheev, Physics Letters
\underline{B394}, 123 (1997),  A.A. Gvozdev, N.V. Mikheev and L.A.
Vassilevskeya, Physics Letters \underline{B410}, 211 (1997).
 \vspace{-7pt}
\item A.N. Ioannisian and G.G. Raffelt, Phys. Rev.
\underline{D55}, 7038 (1997), G.G. Raffelt, Physics Reports
\underline{320}, 319 (1999).
 \vspace{-7pt}
\item R. Shaisultanov, Phys. Rev. Lett. \underline{80}, 1586
(1998), H. Giess and R. Shaisultanov, Phys.Rev. \underline{D62},
073003 (2000).
 \vspace{-7pt}
\item J. Schwinger, Phys. Rev. \underline{82}, 664 (1951);
\underline{93}, 615 (1954).
 \vspace{-7pt}
\item K.S. McFarland and A.C. Melissinos, \lq\lq Quantum Electrodynamics
and Physics of the Vacuum", G. Cantatore ed. AIP Conference
Proceedings \underline{564}, 158 (2001).
 \vspace{-22pt}
\item K. Fujikawa and R.E. Shrock, Phys. Rev. Lett.
\underline{45}, 963 (1980).
 \vspace{-7pt}
\item D.E. Groom et al. (Particle Data Group), Eur. Phys. Jour.
\underline{C15}, 1 (2000).
 \vspace{-7pt}
\item L.A. Ahrens et al., Phys. Rev. \underline{D41}, 3297 (1990).
 \vspace{-7pt}
\item M.C. Gonzales-Garcia, F. Vannucci and J. Castromonte, Phys.
Lett. \underline{B373}, 153 (1996).
 \vspace{-7pt}
\item S. Matsuki and K.Yamamoto, Phys. Lett. \underline{B289}, 194
(1992); F. Vannucci, hep-exp/9911025..
\end{enumerate}

\vfil\eject

\centerline{\underline\bf Figure Captions}

\vspace{.50in}
\begin{figure}[h] \epsfbox {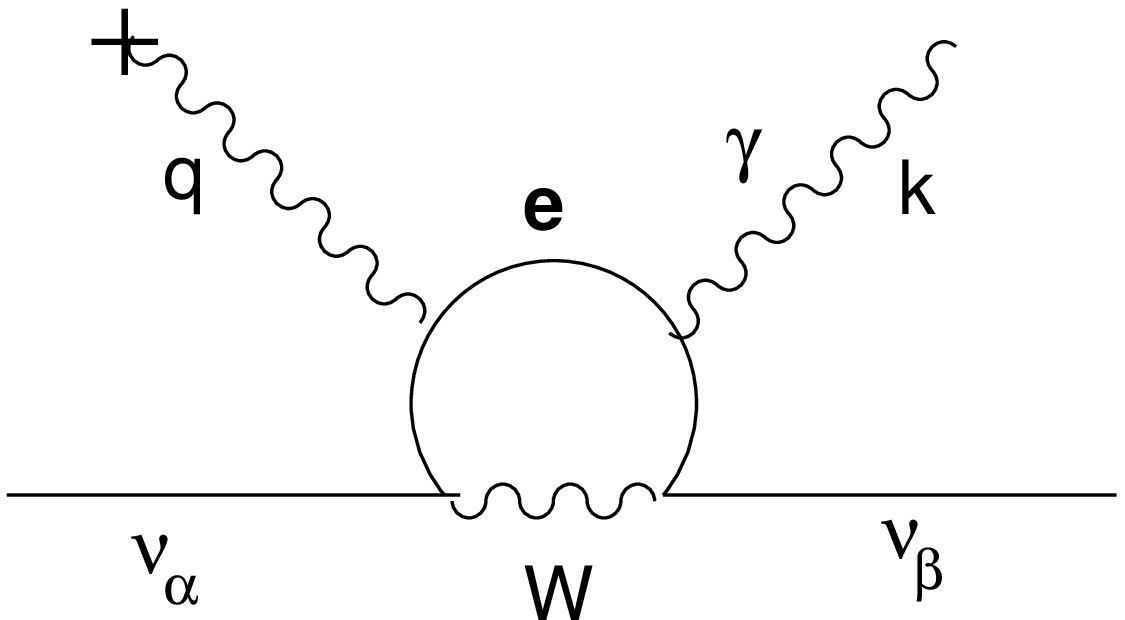}
\label{fig:fig1a.eps}
\end{figure}

\vspace{.50in}
\begin{figure}[h]
 \epsfbox{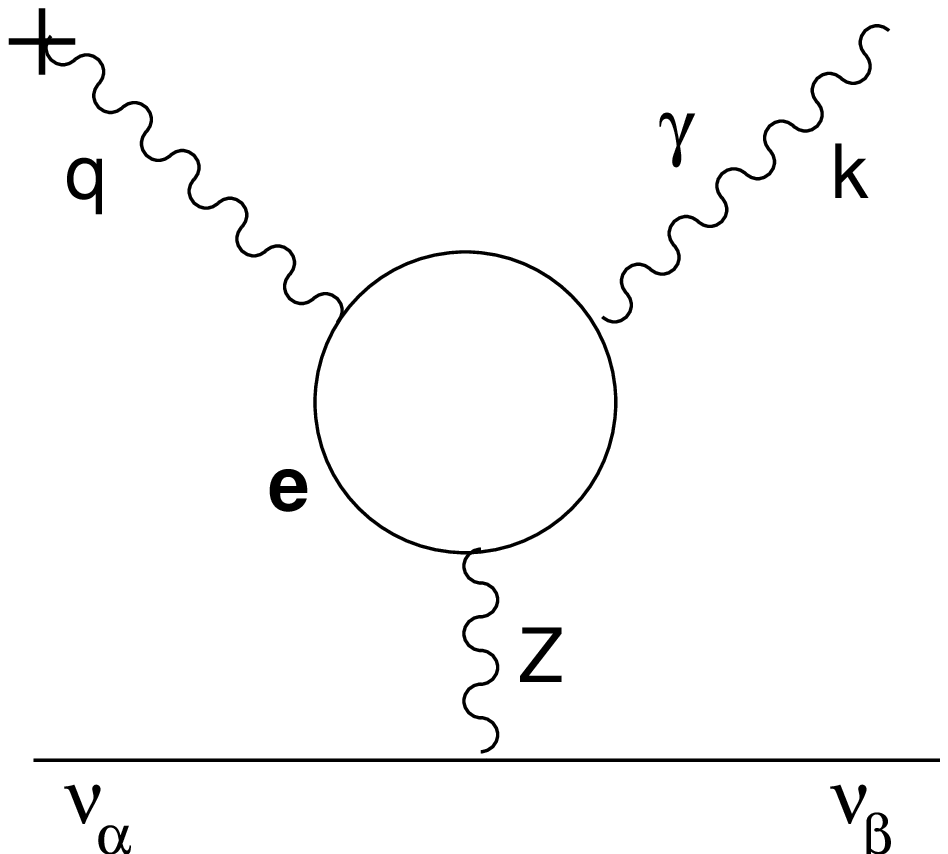} \label{fig:fig1b.eps}
\end{figure}

\begin{description}

\item[Fig.1] The two lowest order diagrams that contribute to
radiative scattering in a magnetic field.

\vfil\eject

\begin{figure}[htbp]
\epsfxsize=\textwidth\epsfbox[30 230 580 530]{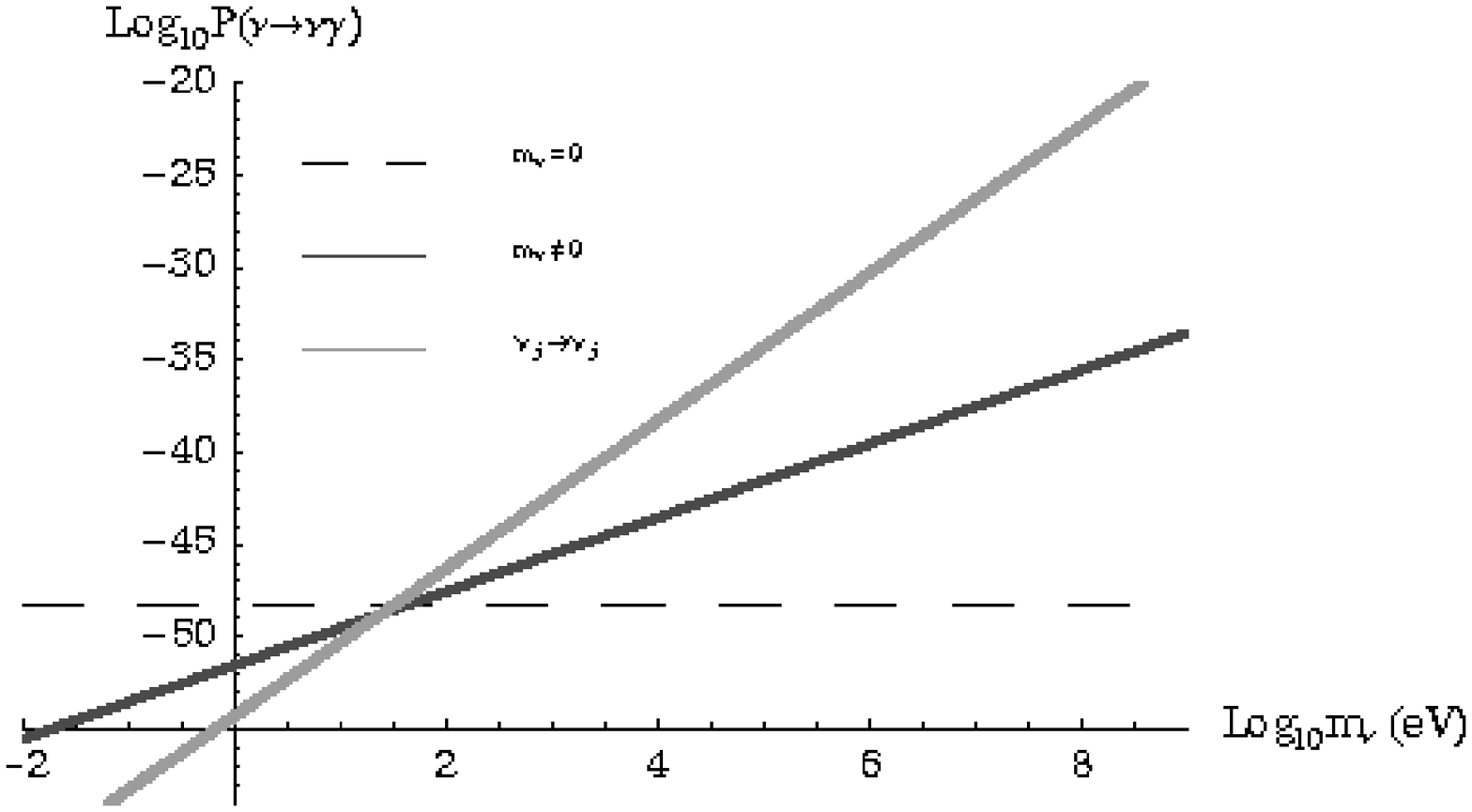}
\label{fig:fig2.eps}
\end{figure}

\vspace{-20pt}

\item[Fig.2] Probability for radiative scattering in a magnetic
field $B$ = 2.2 T and $L = 10\ {\rm m}$ calculated for $E_\nu =
50$ GeV as a function of neutrino mass for the three cases
discussed in the text [Eqs.3-5].  The present limits on neutrino
masses are [16]: $m_{\nu e} < 7\ {\rm eV}, m_{\nu\mu} < 0.17\ {\rm
MeV}, m_{\nu\tau} < 24\ {\rm MeV}$.

\vfil\eject
\begin{figure}[htbp]
\epsfbox{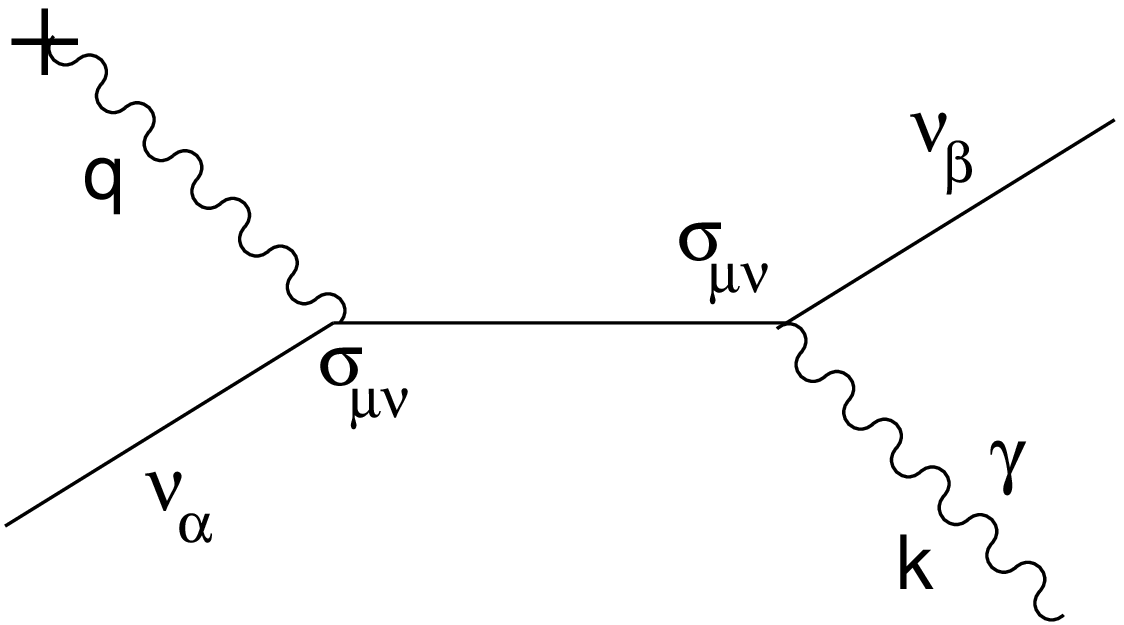} \label{fig:fig3.eps}
\end{figure}

\item[Fig.3] Feynman diagram for radiative scattering in a
magnetic field in the presence of a magnetic moment.

\end{description}

\end{document}